# A SECURITY BASED DATA MINING APPROACH IN DATA GRID

S.Vidhya, S.Karthikeyan

**Abstract -** Grid computing is the next logical step to distributed computing. Main objective of grid computing is an innovative approach to share resources such as CPU usage; memory sharing and software sharing. Data Grids provide transparent access to semantically related data resources in a heterogeneous system. The system incorporates both data mining and grid computing techniques where Grid application reduces the time for sending results to several clients at the same time and Data mining application on computational grids gives fast and sophisticated results to users. In this work, grid based data mining technique is used to do automatic allocation based on probabilistic mining frequent sequence algorithm. It finds frequent sequences for many users at a time with accurate result. It also includes the trust management architecture for trust enhanced security.

______________________________________________

## 1. INTRODUCTION

### 1.1. Grid Computing

A parallel processing architecture in which CPU resources are shared across a network, and all machines function as one large supercomputer, it allows unused CPU capacity in all participating machines to be allocated to one application that is extremely computation intensive and programmed for parallel processing. Grid computing is also called "peer to peer computing" and "distributed computing."

-------------------------------------

- S.Karthikeyan is with Department of Computer Science, SNS College of Technology, Coimbatore, India
- S.Vidhya is with Department of Computer Science, SNS College of Technology, Coimbatore, India

The grid computing gives us yet another way of sharing the computer resource and yields us the maximum benefit at the time and speed efficiency. Grid computing enables multiple applications to share computing infrastructure, resulting in much greater flexibility, cost, power efficiency, performance, scalability and availability at the same time.

### 1.2. Data Grid

A data grid is a grid computing system that deals with the data controlled sharing and management of large amount of distributed data. A Data Grid can include and provide transparent access to semantically related data resources that are different managed by different software systems and are accessible through different protocols and interfaces.

**1.3. Distributed Data Mining**

Distributed data mining deals with the problem of data analysis in environments of distributed computing nodes, and users peer to peer computing is emerging as a new distributed computing for many novel applications that involve exchange of information among a large n umber of peers with little centralized coordination.

**1.4. Data Mining Rule**

SPRINT algorithm for searching the data. Finally it finds result and sends to the server.

In this work, a unified view is provided in which it allows user to use a single query to retrieve all the information transparently from different data sources. The technologies used in this work includes standard for data access over grid, high level data access and semantic data integration Where the high level data access is provided by OGSA-DQP system and semantic data integration by XMAP framework.

Integrating OGSADQP system and XMAP framework has developed the prototype shown in this work.

## 2. DESIGN GOAL

The proposed grid based data mining technique is used to do automatic allocation based on the algorithm Probabilistic mining frequent sequences. This algorithm is used to find frequent sequences in complex databases. It finds frequent sequences for many users at a time with accurate result. Grid application for this application reduces the time for sending results to several clients at the same time. Data mining application on computational grids gives fast and sophisticated results to users.

In this system, data integration architecture needs to combine both the query reformulation and the query processing services. This system offers a wrapper/mediator-based approach to integrate data sources, and adopts the XMAP decentralized mediator approach to handle semantic heterogeneity over data sources. More precisely, the proposed framework is characterized by three core components: a query reformulator engine, a distributed query processor, and a wrapper module as shown in figure 1

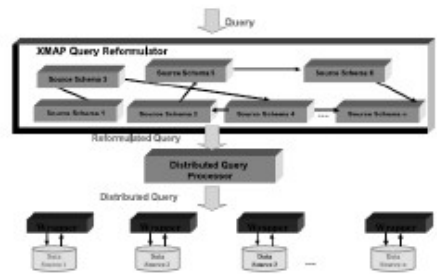

**Figure 1 System Model**

In this figure, XMAP plays the role of the reformulation engine; OGSA-DQP is the distributed query processor. Wrappers are provided by OGSA-DAI. The query from the user is handled by the reformulator engine through the XMAP query reformulation algorithm produces zero, one or more reformulations of the original query.

All the obtained reformulations are then processed by the DQP module that partitions each of such queries in several sub-queries to be executed in parallel. Then, each produced sub-query execution plan is processed. This system is developed with Data mining and Grid computing. Grid computing technology is applied for many applications nowadays. Main objective of grid computing is share resources such as CPU usage, Memory sharing and software sharing.

The proposed grid based data mining technique is used to do automatic allocation based on the algorithm Probabilistic mining frequent sequences. This algorithm is used to find frequent sequences in complex databases. It finds frequent sequences for many users at a time with accurate result. Grid application for this application reduces the time for sending results to several clients at the same time. Data mining application on computational grids gives fast and sophisticated results to users.

## 3. SYSTEM FUNCTIONALITIES

In this work, the grid-computing model includes distributed data mining technique to locate the data present in different location by using query search method. The basic components available in network of the model are: Grid Client, Grid Resource, Grid Scheduler, and Distributed Data Grid.

A unified view is provided in which it allows user to use a single query to retrieve all the information transparently from different data sources. The technologies used in this work includes standard for data access over grid, high level data access and semantic data integration Where the high level data access is provided by OGSA-DQP system and semantic data integration by XMAP framework.

Client always sends queries to scheduler. The scheduler passes those queries to the related resources based on users' requirement. The resources connect with the data grids. The scheduler must know about the resources that are connected with each data grid. Based on this, the scheduler allocates correct resource to the respective query made by the client.

### 3.1. Client Requests the Job to Server
Client module have request ion of water classifications. Next the client sends the subject code and the server to find the particular staff.

### 3.2. Find Suitable Resource Provider
Initially all the gridlets should be connected with the server. Server checks for the minimum CPU utilization time of the connected gridlets and then it allocates the job to the particular gridlet.

### 3.3. Job Monitoring
Server will monitor all the jobs of the gridlet and at the same time it will track the CPU utilization time.

### 3.4. Result Aggregation

Next step is to aggregate the result to produce the final output of the given work by using the result aggregator module.

### 3.5. Send Result to the Specific Client

The server will finally send the aggregate result to the specific client.

### 3.6. Data Mining Rule

SPRINT algorithm is applied here for searching the data. Finally it finds result and sends to the server.

## 4. RELATED WORK

To the best of our knowledge, there are only few works designed to provide schema integration in Grids. The most notable ones are Hyper: A Framework for Peer-to-Peer Data Integration on Grids [5]. This work aims at developing Data Grid integration system based on peer-to-peer schema mappings, which resorts on epistemic logic for providing semantics for the mappings among peers. As in other P2P integration systems, the integration is achieved without using any hierarchical structure for establishing mappings among the autonomous peers. Also, we have presented an OGSA-DAI extension defining data services which are capable of integrating any other standard data service in a Data Grid. Hyper is concerned with high quality query answering on complex structured data, and should not be compared with performance-oriented infrastructures dealing with loosely structured information.

The Grid Data Mediation Service (GDMS) is part of the Grid Miner project and uses a wrapper/mediator approach GDMS presents heterogeneous, distributed data sources as one logical virtual data source in the form of an OGSA-DAI service. This integration model follows an approach not based on a hierarchical structure as well; however it focuses on XML data sources and is based on schema-mappings that associate paths in different schemas. The notion of peer to peer semantic mappings [12] appears also in a non-grid setting explained in this work Schema Mediation for Large-Scale Semantic Data Sharing. An earlier version of this work [2] is Service choreography for data integration on the grid.

## 5. CONCLUSION

The development of practical grid computing techniques will have a profound impact on the way data is analyzed. In particular, the possibility of utilizing grid based data mining applications is very appealing to organizations wanting to analyze data distributed across geographically dispersed heterogeneous platforms.

Grid-based data mining would allow companies to distribute compute-intensive analytic processing among different resources. Moreover, it might eventually lead to new integration and automated analysis techniques that would allow companies to mine data where it resides. This is in contrast to the current practice of having to extract and move data into a centralized location for mining processes that are becoming more difficult to conduct due to the fact that data is becoming increasingly geographically dispersed, and because of security and privacy considerations.

**First Vidhya.S** is a Lecturer of Computer Science and Engineering at SNS College of Technology, Anna University. She received a B.E degree in Computer Science and Engineering from Anna University, India in 2006 and M.E degree in Computer Science and Engineering from Anna University, India in 2009.

**Second Karthikeyan.S** is a Lecturer of Computer Science and Engineering at SNS College of Technology, Anna University. He received a B.Tech degree in Information Technology from Anna University, India in 2006 and M.E degree in Computer Science and Engineering from Anna University, India in 2009.